\documentclass[lettersize,conference]{IEEEtran}
\IEEEoverridecommandlockouts

\usepackage{ifpdf}

\ifCLASSINFOpdf
\usepackage{graphicx}
\graphicspath{{../pdf/}{../jpeg/}}
\DeclareGraphicsExtensions{.pdf,.jpeg,.png}
\else
\usepackage{graphicx}
\graphicspath{{../eps/}}
\DeclareGraphicsExtensions{.eps}
\fi

\usepackage{amssymb}
\usepackage[cmex10]{amsmath}
\usepackage{amsthm}
\usepackage{algorithmic}
\usepackage{algorithm}
\usepackage{array}
\usepackage{eqparbox}
\usepackage{url}
\usepackage[draft]{hyperref}
\usepackage{enumerate}
\usepackage[T1]{fontenc}
\usepackage{multirow}
\usepackage{xcolor}
\usepackage{pdflscape}
\usepackage{fancyhdr}
\usepackage{wrapfig}
\usepackage{pifont}
\usepackage{wasysym}
\usepackage{subcaption}
\usepackage{booktabs}
\usepackage{tikz}
\usepackage{pgfplots}
\pgfplotsset{compat=1.16}
\usetikzlibrary{patterns,patterns.meta,shapes}
\usepackage{longtable}
\usepackage{makecell}
\usepackage[acronym]{glossaries}
\makeglossaries

\usepackage{tabularx}

\newtheorem{assumption}{Assumption}

\newcommand{\bseq}{\begin{subequations}}
	\newcommand{\eseq}{\end{subequations}}
\newcommand{\baln}{\begin{align}}
	\newcommand{\ealn}{\end{align}}
\newcommand{\beq}{\begin{equation}}
	\newcommand{\eeq}{\end{equation}}

\newacronym{3GPP}{3GPP}{3rd Generation Partnership Project}
\newacronym{6g}{6G}{Sixth Generation}
\newacronym{AWGN}{AWGN}{additive white Gaussian noise}
\newacronym{CBRS}{CBRS}{Citizens Broadband Radio Service}
\newacronym{CRLB}{CRLB}{Cram\'er-Rao lower bound}
\newacronym{DoD}{DoD}{Department of Defense}
\newacronym{DSA}{DSA}{Dynamic Spectrum Access}
\newacronym{DSO}{DSO}{Dynamic Spectrum Operations}
\newacronym{DSS}{DSS}{Dynamic Spectrum Sharing}
\newacronym{EIRP}{EIRP}{Equivalent Isotropically Radiated Power}
\newacronym{FCC}{FCC}{Federal Communications Commission}
\newacronym{FR1}{FR1}{Frequency Range 1}
\newacronym{FR2}{FR2}{Frequency Range 2}
\newacronym{FR3}{FR3}{Frequency Range 3}
\newacronym{FSS}{FSS}{Fixed Satellite Service}
\newacronym{FWA}{FWA}{Fixed Wireless Access}
\newacronym{GAA}{GAA}{General Authorized Access}
\newacronym{GEO}{GEO}{Geostationary Orbit}
\newacronym{GNSS}{GNSS}{Global Navigation Satellite System}
\newacronym{GPS}{GPS}{Global Positioning System}
\newacronym{ICaSN}{ICaSN}{Integrated Cellular, Sensing, Navigation, and Radiolocation}
\newacronym{IoT}{IoT}{Internet of Things}
\newacronym{ISAC}{ISAC}{Integrated Sensing and Communication}
\newacronym{ISM}{ISM}{Industrial, Scientific and Medical}
\newacronym{JCAS}{JCAS}{Joint Communications and Sensing}
\newacronym{KPI}{KPI}{key performance indicator}
\newacronym{LEO}{LEO}{Low Earth Orbit}
\newacronym{LTE}{LTE}{Long-Term Evolution}
\newacronym{MEC}{MEC}{Multi-Access Edge Computing}
\newacronym{MIMO}{MIMO}{Multiple-Input Multiple-Output}
\newacronym{MINLP}{MINLP}{Mixed-Integer Nonlinear Program}
\newacronym{mmWave}{mmWave}{millimeter wave}
\newacronym{MNO}{MNO}{Mobile Network Operator}
\newacronym{NGSO}{NGSO}{Non-Geostationary Orbit}
\newacronym{NPRM}{NPRM}{Notice of Proposed Rulemaking}
\newacronym{NR}{NR}{New Radio}
\newacronym{NTN}{NTN}{Non-Terrestrial Network}
\newacronym{NTIA}{NTIA}{National Telecommunications and Information Administration}
\newacronym{ofdm}{OFDM}{Orthogonal Frequency Division Multiplexing}
\newacronym{ORAN}{O-RAN}{Open Radio Access Network}
\newacronym{PAL}{PAL}{Priority Access License}
\newacronym{PFD}{PFD}{Power Flux Density}
\newacronym{prb}{PRB}{Physical Resource Block}
\newacronym{QoS}{QoS}{Quality of Service}
\newacronym{rapp}{rApp}{rApp}
\newacronym{RF}{RF}{radio frequency}
\newacronym{RIC}{RIC}{RAN Intelligent Controller}
\newacronym{RMS}{RMS}{root-mean-square}
\newacronym{SAS}{SAS}{Spectrum Access System}
\newacronym{SINR}{SINR}{signal-to-interference-plus-noise ratio}
\newacronym{SNR}{SNR}{signal-to-noise ratio}
\newacronym{THz}{THz}{terahertz}
\newacronym{UAV}{UAV}{Unmanned Aerial Vehicle}
\newacronym{esc}{ESC}{Environmental Sensing Capability}
\newacronym{sas}{SAS}{Spectrum Access System}
\newacronym{rcs}{RCS}{Radar Cross-Section}
\newacronym{pf}{PF}{Proportional Fair}
\newacronym{peb}{PEB}{position error bound}
\newacronym{SCA}{SCA}{successive convex approximation}

\usepackage{soul}
\usepackage{flushend}

\definecolor{matBlue}   {rgb}{0.00,0.45,0.74}
\definecolor{matOrange} {rgb}{0.85,0.33,0.10}
\definecolor{matYellow} {rgb}{0.93,0.69,0.13}
\definecolor{matPurple} {rgb}{0.49,0.18,0.56}
\definecolor{matGreen}  {rgb}{0.47,0.67,0.19}
\definecolor{navGreen}  {rgb}{0.086,0.639,0.165}
\definecolor{starFill}  {rgb}{1.00,0.84,0.00}
\definecolor{matCyan}    {rgb}{0.30,0.75,0.93}

\pgfplotsset{
	set layers=standard,
}

\pgfplotsset{
	qosplot/.style={
		scale only axis=true,                
		width=0.688\linewidth,                 
		height=0.46\linewidth,                
		grid=both,
		grid style={gray!10, line width=0pt},
		label style={font=\scriptsize},
		title style={font=\scriptsize},
		xticklabel style={font=\scriptsize},
		yticklabel style={font=\scriptsize},
		major tick length=2pt,
		ylabel style={font=\scriptsize, yshift=2pt},
		xlabel style={font=\scriptsize, yshift=2pt},
		xlabel near ticks,
		ylabel near ticks,
		xlabel shift=-3pt,
		ylabel shift=-5pt,
		legend cell align=left,
		legend style={
			font=\tiny, draw=black, line width=0.2pt,
			fill=white, fill opacity=0.9, text opacity=1,
			inner sep=1pt, row sep=-1pt,
		},
		every axis plot/.append style={line width=0.9pt, mark size=1.5pt},
	},
	ratey/.style={
		scaled y ticks=base 10:-3,
		yticklabel style={/pgf/number format/fixed,
			/pgf/number format/precision=0},
		ytick scale label code/.code={$\times 10^{3}$},
	},
}

\fancypagestyle{firstpage}{
	\fancyhf{}
	
	\fancyhead[C]{%
		\fbox{\parbox{\dimexpr\textwidth-2\fboxsep-2\fboxrule\relax}{
                \centering
				\footnotesize\normalfont
				This paper has been accepted for publication at IEEE MILCOM 2026. This is the author’s accepted version of the paper.
		}}
	}
}

\begin{document}
	
	
	
	
	\title{Coordinated Spectrum Coexistence Across Heterogeneous Commercial and Federal Services}
	
	\author{
		\IEEEauthorblockN{Minh Dat Nguyen, Paolo Testolina, Pedram Johari, Michele Polese, Tommaso Melodia}
		\IEEEauthorblockA{Institute for Intelligent Networked Systems, Northeastern University, Boston, MA, U.S.A.
			\\\{minhd.nguyen, p.testolina, p.johari, m.polese, melodia\}@northeastern.edu}
		\thanks{This work was partially supported by the U.S. National Science Foundation under Grant CNS-2434081. This effort was also sponsored by the U.S. Government under Other Transaction number
			W15QKN-21-9-5599 between the National Spectrum Consortium (NSC) and the
			Government. The U.S. Government is authorized to reproduce and distribute reprints
			for Governmental purposes notwithstanding any copyright notation herein. The views and conclusions contained herein are those of the authors and should not
			be interpreted as necessarily representing the official policies or endorsements, either
			expressed or implied, of the U.S. Government.}
		\thanks{  
			DISTRIBUTION STATEMENT A. APPROVED FOR PUBLIC RELEASE; DISTRIBUTION IS UNLIMITED}
		
		\vspace{-10mm}
	}
	
	\maketitle
    \thispagestyle{firstpage}
	
	
	\begin{abstract}
		Future wireless networks are expected to support the coexistence of cellular communications, \gls{RF} sensing, radionavigation, and radiolocation radar---among others---over congested federal sub-6\,GHz spectrum under heterogeneous \gls{QoS} requirements, 
		\textcolor{black}{driven by escalating cellular traffic demand, the proliferation of 6G sensing and positioning services, and regulatory pressure to repurpose federal bands without displacing incumbents.}
		We develop a unified framework in which all four services dynamically share a common \gls{prb} pool under centralized coordination, formulating weighted cellular sum-rate maximization subject to duty-cycle, orthogonality, sensing \gls{SNR}, and Cram\'er--Rao-based positioning constraints. The resulting \gls{MINLP} is solved by alternating optimization across \gls{prb} assignment, scheduling, and successive convex approximation for power allocation, complemented by a low-complexity \gls{QoS}-aware greedy heuristic. Site-specific ray-tracing simulations on the BostonTwin urban digital twin show substantial gains in spectrum efficiency and cellular throughput while strictly meeting sensing and positioning \gls{QoS}, establishing coordinated multi-service sharing as a viable architecture for cellular-federal-radar coexistence in next-generation military and civilian networks.
	\end{abstract}
	
	
	\glsresetall
	\glsunset{NR} 
	\glsunset{3GPP} 
	\glsunset{FCC} 
	\glsunset{NTIA} 

	
	\vspace{-3mm}
	\section{Introduction}
	\label{sec:intro}
	
	The design and standardization of \gls{6g} cellular systems is creating increasing interest in understanding spectrum coexistence mechanisms in multiple sub-6\,GHz bands. This includes assessments of whether sharing is possible across commercial cellular deployments and incumbent federal services (high-power radars, radionavigation infrastructure, and sensing systems)~\cite{SS:Sharingband25}, each governed by distinct \gls{QoS} requirements and regulatory duty-cycle constraints~\cite{fcc_sub6,itu_m1313_band2000}.
	
	
	Static frequency partitioning leaves spectrum idle during off-duty intervals of intermittent services while failing to meet cellular peak-throughput demands~\cite{FCC2002SpectrumPolicy}. \gls{CBRS} demonstrated that three-tier dynamic sharing between satellite downlink, Navy radars, and \gls{LTE} or \gls{NR} carriers is feasible, but only with limited temporal dynamics and integration, preventing performance gains from tight multiplexing of different waveforms. 
	
	\Gls{ISAC} has emerged as a leading paradigm for joint spectrum utilization~\cite{Liu2022Survey}. Early radar-cellular coexistence work~\cite{Saruthirathanaworakun2012JSAC} provided an analytical model for opportunistic cellular access during radar off-duty intervals but is restricted to a two-service model without subcarrier-level optimization or positioning constraints. Subsequent \gls{ISAC} research has advanced joint subcarrier-power allocation for multicarrier and dual-function radar-communication systems~\cite{Wang2021TSP}, 
	while a parallel thread has examined sensing-communication trade-offs~\cite{Dong2023TWC} and joint sensing-localization energy minimization~\cite{Zhang2024JSAC}.
	
	Despite these advances, two critical gaps remain. \emph{First}, \emph{radionavigation}, a co-primary occupant mandated by ITU-R and national regulators~\cite{itu_m1313_band2000} and present in nearly every contested federal band (e.g., aeronautical surveillance and maritime navigation radars), has not been incorporated as an explicit coexisting service with positioning-accuracy constraints in any prior resource allocation framework. 
	Most \gls{ISAC} frameworks adopt a two-service model (communication + sensing/radar) and treat localization as a by-product of \gls{ISAC} waveform design at the UE level, rather than as an independent service with its own infrastructure, users, and regulatory constraints. Navigation accuracy depends jointly on the received signal quality (\gls{SINR}) and the geometric diversity of the available infrastructure anchors, naturally captured by a \gls{CRLB}-based ranging model and a Fisher information matrix-based \gls{peb}~\cite{Shen2010FundamentalLimits}. \emph{Second}, the joint modeling of heterogeneous duty cycles across different service types within a unified coordination architecture has not been addressed.
	

	
	\textcolor{black}{To bridge these gaps, this paper proposes a \emph{coordinated multi-service spectrum sharing} framework, in which cellular, sensing, navigation, and radiolocation share a unified \gls{prb} resource pool. 
		The formulation is not limited to these four services; the resource grid and constraint structure extend naturally to additional service classes (e.g., passive radio astronomy, satellite earth-station receivers). A centralized coordinator is instantiated as an rApp within the \gls{ORAN} Non-RT \gls{RIC}, as it operates over policy-level horizons (seconds to minutes) matching the optimization timescale and has access to RAN-wide telemetry and external data. 
		In this regard, we further assume an \emph{extended coordination plane} in which sensing, radionavigation, and radar nodes expose minimal telemetry (activity schedules, target/anchor lists, \gls{QoS} thresholds) and accept resource grants via O1/E2-like interfaces, a system-design assumption aligned with the broader \gls{DSO} vision for contested federal bands~\cite{dod_emss_2020}, requiring standardization beyond current \gls{ORAN} scope. 
		The framework jointly optimizes long-term \gls{prb} allocation and transmit-power allocation, maximizing the weighted cellular sum-rate subject to service-specific \gls{QoS}: minimum sensing \gls{SNR} and a \gls{CRLB}-based \gls{peb} for radionavigation.} 
	
	The contributions of this paper are as follows:
	\begin{enumerate}
		\item \textbf{Multi-service system model with radionavigation.} We develop the first unified resource allocation framework to incorporate radionavigation as a co-primary coexisting service, characterized by a \gls{CRLB}-based ranging model and a Fisher information matrix-based PEB.
		\item \textbf{Heterogeneous duty-cycle model with dynamic \gls{prb} allocation.} We introduce a generalized activity model capturing heterogeneous duty cycles across all four service types within a unified scheduling framework, coupled with dynamic service-level \gls{prb} assignment over the full spectrum grid.
		\item \textbf{\gls{MINLP} formulation and decomposition.} 
		We formulate the problem as an \gls{MINLP}, establish NP-hardness via reduction from 0/1 knapsack, and develop a three-stage alternating optimization with monotone-convergence guarantees, complemented by a \gls{QoS}-aware greedy algorithm that is linear in \glspl{prb} and time slots.
		\item \textbf{Site-specific evaluation.} We build an evaluation platform on the BostonTwin Sionna-RT scene~\cite{SS:ModelRealistic24}, calibrated to FCC/NTIA frequency assignments in the 2.7--3.7\,GHz federal-band cluster~\cite{ntia_3100_3700} and 3GPP NR resource structure~\cite{3gpp.38.214}, and quantify the cost of \gls{QoS} enforcement under realistic urban propagation.
	\end{enumerate}
	
	\vspace{-2mm}
	\section{System Model and Problem Formulation}
	\label{sec:sysmodel}
	\vspace{-1mm}
	
	\subsection{Network Architecture and Spectrum Coordination}
	\label{subsec:network_model}
	\vspace{-1mm}
	\begin{figure}[t!]
		\centering
		\includegraphics[width=0.92\linewidth]{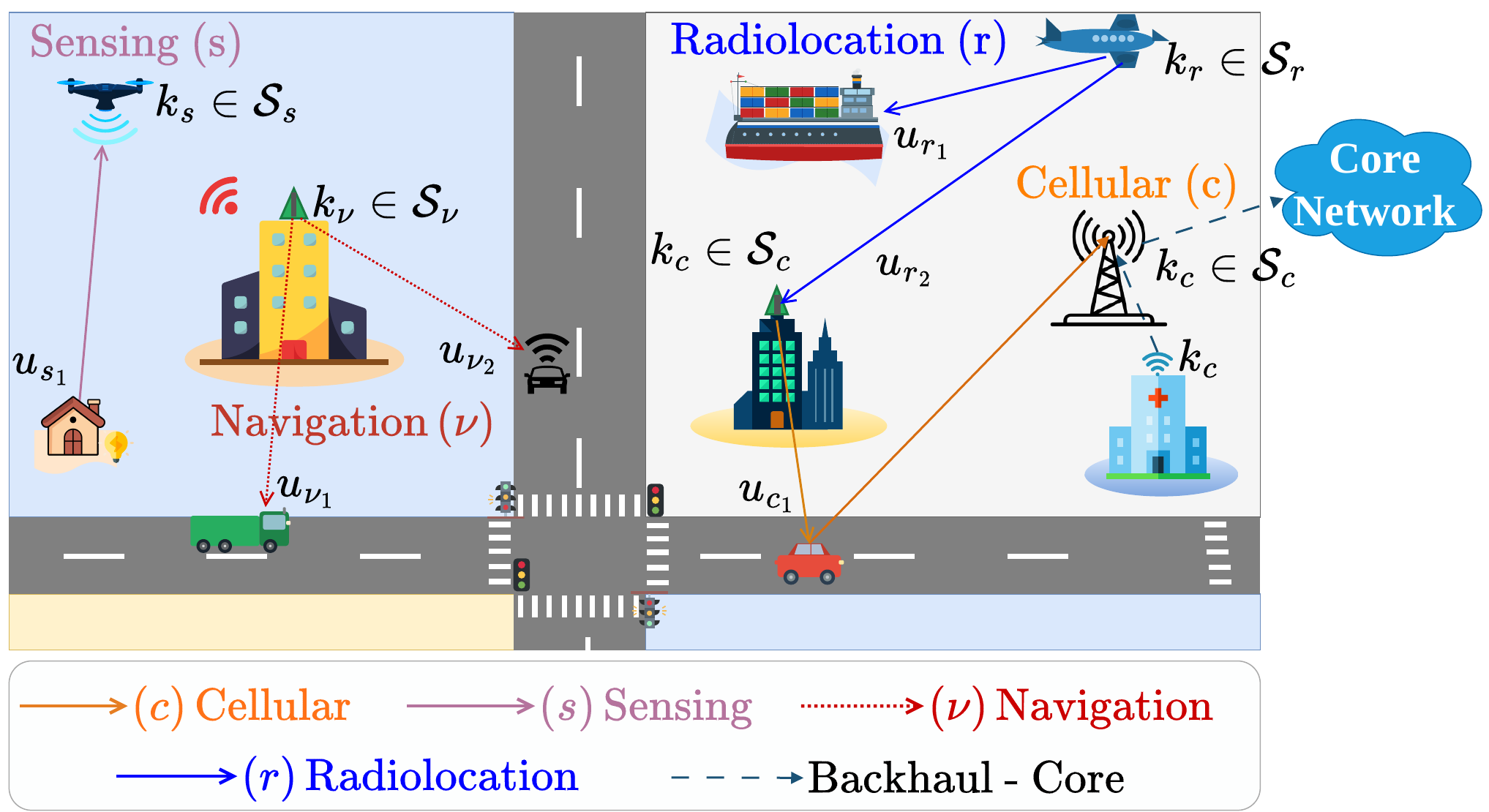}
		\caption{Multi-service network architecture.}
	\label{fig:system_model}
	\vspace{-5mm}
\end{figure}

\textcolor{black}{We consider a multi-service wireless network} supporting the service set $\mathcal{I}\triangleq\{c,s,\nu,r\}$, where $c$, $s$, $\nu$, $r$ denote cellular, \gls{RF} sensing, radionavigation, and radiolocation radar, respectively (Fig.~\ref{fig:system_model}). 
\textcolor{black}{Each service $i$ has transmitters indexed by $k \in \mathcal{S}_i$}  and endpoints: cellular UEs $\mathcal{U}_c$ associated with gNBs by a strongest-power rule; navigation users $\mathcal{U}_\nu$ each connected to a visible anchor set $\mathcal{S}_\nu(u)\subseteq\mathcal{S}_\nu$; sensing/radar nodes monitoring target sets $\mathcal{M}_{i,k}$ for $i\in\{s,r\}$. 

Time is slotted, $t\in\mathcal{T}\triangleq\{1,\ldots,T\}$, with slot duration $\tau$. 
\textcolor{black}{A slot aggregates one or more \gls{ofdm} symbols at the chosen numerology (e.g., $\tau\!=\!1$\,ms with $\Delta f\!=\!30$\,kHz corresponds to a $14$-symbol slot). 
	For non-cellular services, $\tau$ represents the minimum coordination granularity at which resource grants are updated; it does not require those services to use \gls{ofdm} waveforms internally. 
}

We adopt a \emph{generalized \gls{prb} abstraction}: total bandwidth $B$ is divided into $N_{\mathrm{PRB}}$ elementary units of width $B_{\mathrm{rb}}=B/N_{\mathrm{PRB}}$, matching the \gls{3GPP} \gls{NR} \gls{prb} definition for cellular ($B_{\mathrm{rb}}=12\Delta f$, $\Delta f\!\in\!\{15,30,60,120\}$\,kHz) and the minimum allocable spectral chunk for the other services. Let $\mathcal{N}\triangleq\{1,\ldots,N_{\mathrm{PRB}}\}$. Large-scale propagation parameters are quasi-static within the horizon, 
\textcolor{black}{while small-scale fading varies independently across PRBs and slots and is described in detail in the following section.}   

\vspace{-2mm}
\subsection{Resource Allocation Variables}
\label{subsec:variables}
\vspace{-1mm}

Let $O_{i,k}(t)\in\{0,1\}$ denote 
\textcolor{black}{the activity of transmitter $k\in\mathcal{S}_i$}, with the duty-cycle pre-condition $\frac{1}{T}\sum_t O_{i,k}(t)\le\delta_i$ enforced \emph{by external protocol design} (radar pulse repetition intervals, navigation beacon periodicity) and treated as a known input rather than an optimization variable; cellular activity is scheduler-controlled with $\delta_c\!\approx\!1$. 

Two binary variables encode allocation. The grant variable $x_i^{(n)}(t)\!\in\!\{0,1\}$ assigns \gls{prb} $n$ to service $i$; the scheduling variable $a_{i,k,u}^{(n)}(t)\!\in\!\{0,1\}$ allocates \gls{prb} $n$ from transmitter $k$ to endpoint $u$. 
Inter-service exclusivity and intra-service OFDMA orthogonality require, respectively,
\vspace{-1mm}
\begin{align}
	& \sum_{i\in\mathcal{I}}x_i^{(n)}(t)\le 1, \quad \forall n,t, \label{eq:exclusive_new}\\
	& \sum_{u\in\mathcal{U}_i} a_{i,k,u}^{(n)}(t)\le x_i^{(n)}(t)\,O_{i,k}(t),\;\forall i,k,n,t. \label{eq:ofdma_new}
\end{align}
\vspace{-3mm}

\noindent
\textcolor{black}{Eq.~\eqref{eq:ofdma_new} couples the two binary variables: transmitter $k$ of service $i$ may schedule endpoint $u$ on PRB $n$ only if (i) PRB $n$ has been granted to service $i$ ($x^{(n)}_i\!=\!1$) and (ii) transmitter $k$ is active in slot $t$ ($O_{i,k}(t)\!=\!1$).}

The \gls{prb} usage indicator $z_{i,k}^{(n)}(t)\triangleq\sum_{u}a_{i,k,u}^{(n)}(t)\!\in\!\{0,1\}$ follows from~\eqref{eq:ofdma_new}. Power allocation $p_{i,k}^{(n)}(t)\!\ge\!0$ satisfies
\vspace{-2mm}
\begin{align}
	& \textstyle\sum_n p_{i,k}^{(n)}(t)\le P_{i}^{\max}\,O_{i,k}(t),\;\forall i,k,t, \label{eq:power_budget_new}\\
	& 0\le p_{i,k}^{(n)}(t)\le P_{i}^{(n),\max}\,z_{i,k}^{(n)}(t),\;\forall i,k,n,t, \label{eq:peak_power_new}
\end{align}
\textcolor{black}{where $P^{\max}_{i}$ is the total transmit power budget of transmitter on service $i$, $P^{(n),\max}_{i}$ is the per-PRB peak-power cap.}


\vspace{-2mm}
\subsection{Channel and SINR Models}
\label{subsec:channel_model}
\vspace{-1mm}

For the link from transmitter $\ell\in\mathcal{S}_i$ to receiver $u$ at distance $d_{i,\ell,u}$, path loss is $L_{i,\ell,u}=L_{0,i}+10\alpha_i\log_{10}(d_{i,\ell,u}/d_0)+\xi_{i,\ell,u}$, where $L_{0,i}$ is the reference path loss at the reference distance $d_0$ for service $i$, $\alpha_i$ is the path-loss exponent, and $\xi_{i,\ell,u}\!\sim\!\mathcal{N}(0,\sigma_{\xi,i}^2)$ is a log-normal shadow-fading term. The effective channel power gain on \gls{prb} $n$ is
$ g_{i,\ell,u}^{(n)}(t)=G_{i,\ell,u}\,|h_{i,\ell,u}^{(n)}(t)|^2\,10^{-L_{i,\ell,u}/10}, $ 
with $G_{i,\ell,u}$ the antenna gain and $h_{i,\ell,u}^{(n)}$ a Rayleigh/Rician fading coefficient. For endpoint $u$ served by $k$ on \gls{prb} $n$,
\vspace{-3mm}
\begin{equation}
\mathrm{SINR}_{i,k,u}^{(n)}(t)=\frac{p_{i,k}^{(n)}(t)\,g_{i,k,u}^{(n)}(t)}{I_{i,u}^{(n)}(t)+N_0 B_{\mathrm{rb}}},
\label{eq:sinr_new}
\end{equation}
\vspace{-3mm}

\noindent
where $N_0$ is the one-sided thermal-noise power spectral density at the receiver, so that $N_0 B_{\mathrm{rb}}$ is the noise power per \gls{prb}. The intra-service interference (universal frequency reuse within a service) is
$	I_{i,u}^{(n)}(t)\!=\!\!\!\sum_{\ell\in\mathcal{S}_i,\ell\neq k}\!p_{i,\ell}^{(n)}(t)\,g_{i,\ell,u}^{(n)}(t)\,z_{i,\ell}^{(n)}(t).$

\subsection{Service-Specific QoS Models}
\label{subsec:qos_models}

\textbf{Cellular throughput.} The aggregate throughput of $u{\in}\mathcal{U}_c$ is
\begin{equation}
R_u\triangleq\sum_{t,n} a_{c,k_c(u),u}^{(n)}(t)\,B_{\mathrm{rb}}\log_2\!\bigl(1+\mathrm{SINR}_{c,k_c(u),u}^{(n)}(t)\bigr),
\label{eq:R_u_new}
\end{equation}
\vspace{-4mm}

\noindent
where $k_c(u)\in\mathcal{S}_c$ denotes the gNB serving cellular user $u$.

\textbf{Sensing/radar \gls{SNR}.} Adopting a monostatic topology and the radar range equation~\cite{Skolnik2008Radar}, we define the (slot-independent under quasi-static range/\gls{rcs}) echo-gain coefficient as
$ \eta_{i,k,m}^{(n)}\triangleq\frac{(G_{i,k}^{\mathrm{tx}})^2\lambda_i^2\sigma_{\mathrm{RCS},m}}{(4\pi)^3 d_{k,m}^4 L_{\mathrm{sys},i}},\; \forall i\in\{s,r\},$
where $G_{i,k}^{\mathrm{tx}}$ denotes the transmit antenna gain of transmitter $k$ in service $i$, $\lambda_i$ is the carrier wavelength, $\sigma_{\mathrm{RCS},m}$ is the \gls{rcs} of target $m$, $d_{k,m}$ is the one-way range between transmitter $k$ and target $m$, and $L_{\mathrm{sys},i}\!\ge\!1$ aggregates implementation losses (cabling, mismatch, and processing). The echo power is therefore linear in the per-\gls{prb} transmit power $p_{i,k}^{(n)}(t)$: $P_{\mathrm{echo},k,m}^{(n)}(t)=\eta_{i,k,m}^{(n)}\,p_{i,k}^{(n)}(t)$. 
Under equal-weight integration across active \glspl{prb} (a robust low-complexity combiner~\cite{Skolnik2008Radar}), the effective sensing \gls{SNR} for target $m\!\in\!\mathcal{M}_{i,k}$ when the sensing-mode indicator $O_{i,k}(t)\!=\!1$ is
$	\mathrm{SINR}_{i,k,m}^{\mathrm{sen}}(t)=\frac{\sum_n z_{i,k}^{(n)}(t)\,\eta_{i,k,m}^{(n)}\,p_{i,k}^{(n)}(t)}{\sum_n z_{i,k}^{(n)}(t)\bigl(I_{i,k}^{(n)}(t)+N_0 B_{\mathrm{rb}}\bigr)}. $
Therefore, the detection \gls{QoS} is  
\vspace{-1mm}
\begin{eqnarray}
\mathrm{SINR}_{i,k,m}^{\mathrm{sen}}(t)\!\ge\!\gamma_i^{\mathrm{sen}},
\label{con:sensing}
\end{eqnarray}
\vspace{-5mm}

\noindent
where the threshold $\gamma_i^{\mathrm{sen}}$ is calibrated to the desired probability of false alarm.

\textbf{Navigation accuracy.} For anchor $k\!\in\!\mathcal{S}_\nu(u)$, the \gls{prb}-aggregated ranging \gls{SINR} and effective bandwidth are~\cite{Shen2010FundamentalLimits}:
$	\Gamma_{u,k}^{\mathrm{nav}}(t) = \frac{\sum_n a_{\nu,k,u}^{(n)}(t)p_{\nu,k}^{(n)}(t)g_{\nu,k,u}^{(n)}(t)}{\sum_n a_{\nu,k,u}^{(n)}(t)\bigl(I_{\nu,u}^{(n)}(t)+N_0 B_{\mathrm{rb}}\bigr)}, 
B_{\mathrm{eff},u,k}(t) = B_{\mathrm{rb}}\!\sum_n a_{\nu,k,u}^{(n)}(t). $
The \gls{CRLB}-inspired ranging variance is $\sigma_{u,k}^2(t)=c_{0}^2/[8\pi^2 B_{\mathrm{eff},u,k}^2(t)(\Gamma_{u,k}^{\mathrm{nav}}(t)+\varepsilon_0)]$~\cite{Shen2010FundamentalLimits,Win2007Position}, where $c_{0}$ denotes the speed of light and $B_{\mathrm{eff},u,k}(t)$ is a proxy for the \gls{RMS} signal bandwidth and $\varepsilon_0\!>\!0$ is a regularization constant ($\varepsilon_0\!=\!10^{-3}$ in our experiments). The proxy $B_{\mathrm{eff},u,k}(t)$ is exact under uniform spectral allocation and \emph{upper-bounds} the true \gls{CRLB}-based variance otherwise (since contiguous allocations minimize the \gls{RMS} bandwidth for a given $|\sum_n a^{(n)}_{\nu,k,u}(t)|$); the resulting PEB constraint is therefore conservative. Defining $\beta_{u,k}^{(n)}(t)\triangleq 8\pi^2 B_{\mathrm{rb}}^2 g_{\nu,k,u}^{(n)}(t)/c_{0}^2$, the 2D Fisher information matrix
is
$	\mathbf{J}_u(t)=\sum_{k\in\mathcal{S}_\nu(u)}\sigma_{u,k}^{-2}(t)\,\mathbf{v}_{k,u}\mathbf{v}_{k,u}^{\mathsf{T}}, $
where $\mathbf{v}_{k,u}=[\cos\theta_{k,u},\sin\theta_{k,u}]^{\mathsf{T}}$ is the unit line-of-sight direction vector from anchor $k$ to user $u$ in the 2D plane, with $\theta_{k,u}$ the corresponding azimuth angle. The PEB is $\mathrm{PEB}_u(t){=}\sqrt{\mathrm{tr}(\mathbf{J}_u^{-1}(t))}$~\cite{Win2007Position}, and the navigation \gls{QoS} is
\vspace{-3mm}
\begin{equation}
\frac{1}{T}\sum_{t=1}^{T}\mathrm{PEB}_u(t)\le\epsilon_{\max},\quad\forall u\in\mathcal{U}_\nu,
\label{eq:peb_constraint_new}
\end{equation}
\vspace{-4mm}

\noindent
where $\epsilon_{\max}\!>\!0$ is time-averaged positioning-error tolerance. 



\vspace{-2mm}
\subsection{Joint Optimization Problem}
\label{subsec:problem}
\vspace{-1mm}

The coordinated multi-service resource allocation problem, with the objective of maximizing the weighted cellular sum rate, can be formulated as:
\vspace{-2.5mm}
\begin{subequations}\label{P:main}
\begin{align}
	\textbf{(P):}\;\max_{\{\mathbf{x}, \mathbf{a}, \mathbf{p}\}}\quad & \sum_{u\in\mathcal{U}_c} w_u R_u \tag{\ref{P:main}}\\
	\text{s.t.}\quad & \eqref{eq:exclusive_new}\text{--}\eqref{eq:peak_power_new}, \eqref{con:sensing}, \eqref{eq:peb_constraint_new}, \nonumber\\
	& x_i^{(n)}(t),\,a_{i,k,u}^{(n)}(t)\in\{0,1\}, \label{binary_contr}
\end{align}
\end{subequations}
\vspace{-6mm}

\noindent
where $\mathbf{x} = \{x_i^{(n)}(t), \forall i,n,t\}$, $\mathbf{a} = \{a_{i,k,u}^{(n)}(t), \forall i,k,u,n,t\}$, $\mathbf{p} = \{p_{i,k}^{(n)}(t), \forall i,k,n,t\}$, and $w_u$ is the priority weight of cellular user $u$.

Problem \textbf{(P)} prioritizes cellular throughput while enforcing sensing and positioning as hard \gls{QoS} constraints, reflecting the asymmetric urgency of these services in federal-band sharing.
\textbf{(P)} is a \gls{MINLP} due to binary variables and nonconcave \gls{SINR} coupling, and is NP-hard: restricting to $\mathcal{I}\!=\!\{c\}$, $T\!=\!1$, and fixed equal power $p_{c,k}^{(n)}\!=\!\bar{p}$ reduces \textbf{(P)} to maximizing $\sum_n a_{c,k,u}^{(n)} R_n$ subject to $\sum_n a_{c,k,u}^{(n)}\bar{p}\!\le\!P_c^{\max}$, where $R_n\!\triangleq\!w_u B_{\mathrm{rb}}\log_2(1+\bar{p}\,g_{c,k,u}^{(n)}/(N_0 B_{\mathrm{rb}}))$ is the weighted per-\gls{prb} rate, which is exactly the
0/1 knapsack problem, known to be NP-hard~\cite{Martello1990Knapsack}.

\vspace{-2mm}
\section{Decomposition-Based Optimization}
\label{sec:solution}
\vspace{-1.5mm}
Due to its NP-hard nature, we decompose problem~\textbf{(P)} into three subproblems: \gls{prb} assignment, scheduling, and power allocation. These are solved iteratively in the \emph{Alternating Multi-Service Resource Allocation} (AMRA) algorithm, an alternating optimization (AO) loop until convergence. We describe how to solve each subproblem in the following.

\textbf{\gls{QoS} surrogates:} At a current power iterate $\bar{\mathbf{p}}$, define the receiver-side intra-service interference $\bar{I}_{i,k}^{(n)}(t)\!=\!\sum_{\ell\neq k}\bar{p}_{i,\ell}^{(n)}g_{i,\ell,k}^{(n)}z_{i,\ell}^{(n)}$ for $i\!\in\!\{s,r\}$ and $\bar{I}_{\nu,u}^{(n)}$ analogously for navigation. The sensing threshold and navigation information aggregate (both affine in $\mathbf{p}$) are
\vspace{-2mm}
\begin{align}
\Gamma_{i,k,m}^{\mathrm{sen}}(t) \triangleq \gamma_i^{\mathrm{sen}}\sum_n z_{i,k}^{(n)}(t)\bigl(\bar{I}_{i,k}^{(n)}(t)+N_0 B_{\mathrm{rb}}\bigr), \qquad\quad \label{eq:Gamma_sen_def}
\end{align}
\vspace{-6mm}
\begin{align}
Q_u(t;\bar{\mathbf{p}}) \!=\! \!\!\!\sum_{k\in\mathcal{S}_\nu(u)}\!\!\sum_n \frac{a_{\nu,k,u}^{(n)}(t)\,8\pi^2 B_{\mathrm{rb}}^2 g_{\nu,k,u}^{(n)}(t)\,p_{\nu,k}^{(n)}(t)}{c_{0}^2\bigl(\bar{I}_{\nu,u}^{(n)}(t)+N_0 B_{\mathrm{rb}}\bigr)}. \label{eq:nav_info}
\end{align}
\vspace{-4mm}
\subsubsection*{Stage 1: \gls{prb} Assignment}
With $\{\mathbf{a},\mathbf{p}\}$ fixed, we solve
\vspace{-2mm}
\begin{subequations}\label{P1}
\begin{align}
	\textbf{(P1):}\;\max_{\mathbf{x}}\;&\; \sum_{n,t,i}x_i^{(n)}(t)\,\hat{R}_i^{(n)}(t)\tag{\ref{P1}}\\
	\text{s.t.}\;&\; \eqref{eq:exclusive_new}, \eqref{binary_contr},\;x_i^{(n)}(t)\ge z_{i,k}^{(n)}(t), \;\forall i,k,n,t,
\end{align}
\end{subequations}
\vspace{-6mm}

\noindent
where the per-service \gls{prb} benefit is
$	\hat{R}_c^{(n)}(t) = \!\!\sum_{k,u}\!\!w_u a_{c,k,u}^{(n)}(t)B_{\mathrm{rb}}\log_2\!\bigl(1+\overline{\mathrm{SINR}}_{c,k,u}^{(n)}(t)\bigr),
\hat{R}_i^{(n)}(t) = \mu_i\!\!\sum_{k,m\in\mathcal{M}_{i,k}}\!\!\eta_{i,k,m}^{(n)}\bar{p}_{i,k}^{(n)}(t),\quad \forall i\in\{s,r\}, 
\hat{R}_\nu^{(n)}(t) = \mu_\nu\!\!\sum_{u,k\in\mathcal{S}_\nu(u)}\!\!\beta_{u,k}^{(n)}(t)\bar{p}_{\nu,k}^{(n)}(t), $
all evaluated at $\bar{\mathbf{p}}$, with $\mu_i,\mu_\nu\!>\!0$ service priority weights. The consistency constraint $x_i^{(n)}(t)\!\ge\!z_{i,k}^{(n)}(t)$ ensures the updated $\mathbf{x}$ does not revoke a grant already used by $\mathbf{a}$, preserving feasibility of~\eqref{eq:ofdma_new} across AO iterations; a softer variant that allows revocation with a Stage-2 feasibility-restoration step accelerates convergence at the cost of additional per-iteration work and is left to future investigation. \textbf{(P1)} is a binary LP problem, solvable greedily
or using CVX-Gurobi solver \cite{CVX}. 

\subsubsection*{Stage 2: \gls{prb} Scheduling}
With $\{\mathbf{x},\mathbf{p}\}$ fixed, we solve
\vspace{-2mm}
\begin{subequations}\label{P2}
\begin{align}
	\textbf{(P2):}\;\max_{\mathbf{a}}\;& \sum_{u\in\mathcal{U}_c} w_u R_u \tag{\ref{P2}}\\
	\text{s.t.}\;&\; \eqref{eq:ofdma_new}, \eqref{binary_contr}, \\
	& \!\sum_n z_{i,k}^{(n)}(t)\eta_{i,k,m}^{(n)}\bar{p}_{i,k}^{(n)}(t)\!\ge\!\Gamma_{i,k,m}^{\mathrm{sen}}(t), \label{C:P2_sensing}\\
	& \tfrac{1}{T}\textstyle\sum_t Q_u(t;\bar{\mathbf{p}})\ge Q_u^{\min},\;\forall u\in\mathcal{U}_\nu, \label{C:P2_nav}
\end{align}
\end{subequations}
\vspace{-4mm}

\noindent
where $Q_u^{\min}\!\triangleq\!4/\epsilon_{\max}^2$ (justified in Stage~3). With $\mathbf{p}$ fixed, the rate is constant in $\mathbf{a}$ and the \gls{QoS} surrogates~\eqref{C:P2_sensing}--\eqref{C:P2_nav} are affine in $\mathbf{a}$ via $z_{i,k}^{(n)}(t)\!=\!\sum_u a_{i,k,u}^{(n)}(t)$, so \textbf{(P2)} is a binary LP, solvable greedily or using a solver such as CVX-Gurobi \cite{CVX}.

\subsubsection*{Stage 3: Power Allocation via SCA}
\label{subsubsec:power_sca}
With $\{\mathbf{x},\mathbf{a}\}$ fixed, the rate admits a difference-of-concave (DC) form $r_u^{(n)}(t)\!=\!a_{i,k_c(u),u}^{(n)}(t) B_{\mathrm{rb}}[\log_2(S_u^{(n)}(t)\!+\!I_u^{(n)}(t)\!+\!N_0 B_{\mathrm{rb}})\!-\!\log_2(I_u^{(n)}(t)\!+\!N_0 B_{\mathrm{rb}})]$ with $S_u^{(n)}(t)\!=\!p_{c,k_c(u)}^{(n)}(t)g_{c,k_c(u),u}^{(n)}(t)$. Linearizing the second (concave) term at $\bar{\mathbf{p}}$ via first-order Taylor expansion yields a concave lower bound $\widetilde{R}_u(\bar{\mathbf{p}})\!\le\!R_u$.
Echo power being linear in $\mathbf{p}$, the sensing constraint becomes affine, $\sum_n z_{i,k}^{(n)}(t)\eta_{i,k,m}^{(n)}p_{i,k}^{(n)}(t)\!\ge\!\Gamma_{i,k,m}^{\mathrm{sen}}(t)$.

\vspace{-1mm}
\begin{assumption}[Isotropic anchor geometry]
\label{assump:isotropic}
For every $u\!\in\!\mathcal{U}_\nu$ and $t$, the visible anchor geometry is approximately isotropic, i.e., $\mathbf{J}_u(t)\!\approx\!\frac{1}{2}\mathrm{tr}(\mathbf{J}_u(t))\mathbf{I}_2$.
\end{assumption}

\vspace{-2mm}
Under Assumption~\ref{assump:isotropic}, $\mathrm{tr}(\mathbf{J}_u^{-1})\!=\!4/\mathrm{tr}(\mathbf{J}_u)$, so $\mathrm{PEB}_u(t)\!\le\!\epsilon_{\max}\!\Leftrightarrow\!\mathrm{tr}(\mathbf{J}_u(t))\!\ge\!4/\epsilon_{\max}^2$. Lower-bounding the ranging-variance denominator at $\bar{\mathbf{p}}$ yields the affine sufficient condition $\frac{1}{T}\sum_t Q_u(t;\bar{\mathbf{p}})\!\ge\!Q_u^{\min}\!=\!4/\epsilon_{\max}^2$. Hence, the convex SCA subproblem is formulated as 
\vspace{-2mm}
\begin{subequations}\label{P3_sca}
\begin{align}
	\textbf{(P3.1):}\;\max_{\mathbf{p}}\;& \sum_{u\in\mathcal{U}_c} w_u\,\widetilde{R}_u(\bar{\mathbf{p}})\tag{\ref{P3_sca}}\\
	\text{s.t.}\;&\; \eqref{eq:power_budget_new},\,\eqref{eq:peak_power_new},\;\text{and the affine \gls{QoS} surrogates},\nonumber
\end{align}
\end{subequations}
\vspace{-6mm}

\noindent
which has a concave objective and affine constraints and is solved by the CVX-Mosek/Gurobi solver~\cite{CVX}. 

The joint algorithm is summarized in Alg.~\ref{alg:alter_all}. Let $N_p = S_c NT$ denote the number of power variables. \textit{Stage~1} performs $NT$ greedy \gls{prb} selections at $\mathcal{O}\bigl(NT(S_c U_c + S_s \bar{M}_s + S_r \bar{M}_r + U_\nu \bar{A})\bigr)$, where $\bar{M}_s,\bar{M}_r$ are the average numbers of targets per sensing/radar node and $\bar{A} \triangleq \frac{1}{|\mathcal{U}_\nu|}\sum_{u} |\mathcal{S}_\nu(u)|$ the average visible-anchor count per navigation user. \textit{Stage~2} solves binary assignment at $\mathcal{O}\!\left(\sum_i S_i U_i NT\right)$, and \textit{Stage~3} solves~\textbf{(P3.1)} via interior-point methods at $\mathcal{O}(N_p^{3.5})$ per SCA iteration, requiring $N_{\mathrm{sca}}\!=\!\mathcal{O}(1/\epsilon_{\mathrm{sca}})$ iterations~\cite{book:Convex_Boyd}. Stage~3 dominates when $S_c^2 NT^2\!\gg\!\sum_i U_i$, yielding overall complexity 
\vspace{-1mm}
\begin{equation}
\mathcal{O}\bigl(N_{\mathrm{iter}}\,N_{\mathrm{sca}}\,(S_c NT)^{3.5}\bigr).
\label{eq:complexity_overall}
\end{equation}
\vspace{-5mm}

\noindent
Although \textbf{(P)} is NP-hard, the decomposition is polynomial per iteration.
The cubic scaling in $S_c NT$ nevertheless renders~\eqref{eq:complexity_overall} prohibitive at scale, since the interior-point solver is re-invoked at every SCA iteration of every AO cycle, motivating the low-complexity heuristic of the following section.

\setlength{\dbltextfloatsep}{0pt}
\begin{algorithm}[t!]
\scriptsize
\caption{\small Alternating Multi-Service Resource Allocation (AMRA)}
\label{alg:alter_all}
\begin{algorithmic}[1]
	\STATE \textbf{Input:} $\{g_{i,k,u}^{(n)}\}$, $\{O_{i,k}\}$, $\gamma_i^{\mathrm{sen}}$, $\epsilon_{\max}$, $\{w_u,\mu_i\}$.
	\STATE \textbf{Init:} round-robin $\mathbf{x}^{(0)}$, best-PRB $\mathbf{a}^{(0)}$, equal-power $\mathbf{p}^{(0)}$; if~\eqref{con:sensing} and \eqref{eq:peb_constraint_new} are violated, run a feasibility-recovery boost (allocate additional PRBs/power to deficit services); $\ell\!\leftarrow\!0$.
	\REPEAT
	\STATE Solve \textbf{(P1)} $\Rightarrow \mathbf{x}^{(\ell+1)}$.
	\STATE Solve \textbf{(P2)} at $\bar{\mathbf{p}}\!=\!\mathbf{p}^{(\ell)}$ $\Rightarrow \mathbf{a}^{(\ell+1)}$.
	\STATE Iterate \textbf{(P3.1)} until $|\Delta\text{obj}|\!<\!\epsilon_{\mathrm{SCA}}$ $\Rightarrow \mathbf{p}^{(\ell+1)}$; $\ell\!\leftarrow\!\ell+1$.
	\UNTIL{$|\sum_u w_u R_u^{(\ell)}-\sum_u w_u R_u^{(\ell-1)}|\!<\!\epsilon_{\mathrm{AO}}$.}
	\STATE \textbf{Output:} $\{\mathbf{x}^{(\ell)},\mathbf{a}^{(\ell)},\mathbf{p}^{(\ell)}\}$.
\end{algorithmic}
\end{algorithm}

\vspace{-2mm}
\section{QoS-Aware Greedy Heuristic}
\label{sec:heuristic}
\vspace{-1mm}

We propose the \emph{QoS-Aware Greedy Alternating Allocation (QGAA)} heuristic, which prioritizes sensing/navigation feasibility while opportunistically maximizing weighted cellular throughput. At each slot $t$, QGAA executes four steps.

\textit{Step 1---\gls{QoS} Reservation.} For each violator of~\eqref{con:sensing} and \eqref{eq:peb_constraint_new}, candidate \glspl{prb} are ranked by marginal \gls{QoS} improvement (sensing \gls{SNR} gain $\Delta\mathrm{SINR}_{i,k,m}^{\mathrm{sen}}$ for $i\!\in\!\{s,r\}$ or PEB reduction $-\Delta\mathrm{PEB}_u^{(n)}$ for $i\!=\!\nu$) and reserved greedily until violations clear or resources are exhausted.

\textit{Step 2---Greedy \gls{prb} Assignment.} Each remaining \gls{prb} is awarded to $i^\star\!=\!\arg\max_j \Psi_j^{(n)}(t)$, where
\vspace{-2mm}
\begin{equation}
\Psi_i^{(n)}(t)=\!\!\begin{cases}
	\!\max_{k,u} w_u B_{\mathrm{rb}}\log_2\!\bigl(1\!+\!\overline{\mathrm{SINR}}_{c,k,u}^{(n)}(t)\bigr), & i=c,\\
	\!\max_{k,m}\Delta\mathrm{SINR}_{i,k,m}^{\mathrm{sen}}(t), & i\in\{s,r\},\\
	\!\max_{k,u}\bigl(-\Delta\mathrm{PEB}_u^{(n)}(t)\bigr), & i=\nu.
\end{cases}
\label{eq:service_score}
\end{equation}

\textit{Step 3---Intra-Service Scheduling.} Cellular \glspl{prb} go to the user with the largest weighted-rate gain; sensing/radar \glspl{prb} to the target with the largest \gls{SNR} deficit; navigation \glspl{prb} to anchors serving users with the largest positioning error.

\textit{Step 4---Power Adjustment.} Power is initialized uniformly, $p_{i,k}^{(n)}(t)\!=\!P_{i}^{\max}/\sum_{n'}z_{i,k}^{(n')}(t)$. If \gls{QoS} violations persist, a finite-step water-filling-style repair is applied: at each iteration, identify the cellular \gls{prb} $n^\star$ with the smallest marginal cellular rate $\partial R/\partial p_{c,k}^{(n^\star)}$ and the deficit-service \gls{prb} $n^\dagger$ with the largest \gls{QoS} gain per unit power; transfer a step $\Delta p\!=\!\rho\,p_{c,k}^{(n^\star)}$ ($\rho\!\in\!(0,1]$) from $n^\star$ to $n^\dagger$, subject to peak-power caps~\eqref{eq:peak_power_new}. The procedure terminates when all \gls{QoS} constraints are met, when $p_{c,k}^{(n^\star)}\!=\!0$, or after $K_{\max}$ iterations.

\setlength{\dbltextfloatsep}{0pt}
\begin{algorithm}[t!]
\scriptsize
\caption{\small QoS-Aware Greedy Alternating Allocation (QGAA)}
\label{alg:heuristic}
\begin{algorithmic}[1]
	\STATE \textbf{Input, Init:} $\{\mathbf{x}^{(0)},\mathbf{a}^{(0)},\mathbf{p}^{(0)}\}$ as in Algorithm~\ref{alg:alter_all}.
	\REPEAT
	\FOR{each $t\in\mathcal{T}$}
	\STATE Reserve PRBs for QoS violators (Step 1).
	\STATE For each $n$: assign to $\arg\max_i\Psi_i^{(n)}(t)$ via~\eqref{eq:service_score} (Step 2).
	\STATE Intra-service scheduling (Step 3); greedy power repair (Step 4).
	\ENDFOR
	\UNTIL{$|\Delta\sum_u w_u R_u|\!<\!\epsilon_{\mathrm{QGAA}}$.}
	\STATE \textbf{Output:} $\{\mathbf{x},\mathbf{a},\mathbf{p}\}$.
\end{algorithmic}
\end{algorithm}

The dominant cost is computing~\eqref{eq:service_score} and per-service scheduling, yielding overall complexity $\mathcal{O}\bigl(|\mathcal{N}|T\sum_i|\mathcal{S}_i||\mathcal{U}_i|\bigr)$, linear in \glspl{prb} and slots and substantially below the \gls{MINLP}, making QGAA suitable for real-time deployment as an \gls{ORAN} Near-RT \gls{RIC} xApp.

\vspace{-3mm}
\section{Numerical Results}
\label{sec:result}
\vspace{-1mm}


\begin{table}[t!]
\centering
\caption{Simulation Parameters}
\label{tab:sim_params}
\footnotesize

\textit{General Parameters}\\[3pt]
\begin{tabularx}{\linewidth}{lXr}
	\toprule
	\textbf{Parameter} & \textbf{Symbol} & \textbf{Value} \\
	\midrule
	Network area & $L\times L$ & $1$\,\text{km}$\times$ $1$\,\text{km} \\
	Coordination bandwidth & $B$ & $100$\,MHz \\
	Subcarrier spacing & $\Delta f$ & $30$\,kHz \\
	Time slots / slot duration & $T,\tau$ & $10$, $1$\,ms \\
	Noise PSD & $N_0$ & $-174$\,dBm/Hz \\
	Path-loss exp. / shadowing & $\alpha_{i},\sigma_{\xi,i}^2$ & $2.5$--$3.5$, $6$\,dB \\
	Receiver height & $H_{\mathrm{UE}}$ & $1$\,m \\
	Total / per-PRB power & $P^{\max},P_{\mathrm{PRB}}^{\max}$ & $10$\,W, $1$\,W \\
	Sensing SNR threshold & $\gamma^{\mathrm{sen}}$ & $3\;(\approx4.8\,\text{dB})$ \\
	Navigation PEB threshold & $\epsilon_{\max}$ & $2.83$\,m \\
	Navigation info threshold & $Q_u^{\min}$ & $4/\epsilon_{\max}^2\!=\!0.5$ \\
	Duty cycles ($c,s,\nu,r$) & $\delta_i$ & $(1,0.2,0.1,0.1)$ \\
	Multiplied network density & $\zeta$ & $[1,3]$ \\
	\bottomrule
\end{tabularx}

\vspace{6pt}
\textit{Per-Service Parameters}\\[3pt]
\begin{tabularx}{\linewidth}{lcccc}
	\toprule
	\textbf{Service} & 
	\makecell{\textbf{Density} \\ $\zeta_{\sf tx}$/$\zeta_{\sf ue}$} & 
	\makecell{\textbf{TX Height} \\ $H_{\sf TX}$ (m)} & 
	\makecell{\textbf{Freq.} \\ (GHz)} & 
	\makecell{\textbf{TX Power} \\ (dBm)} \\
	\midrule
	Cellular ($c$)    & $60/100/\text{km}^2$ & $25$ & $2$--$4$     & $46$ \\
	Sensing ($s$)     & $30/50/\text{km}^2$ & $30$  & $3.1$--$3.7$ & $50$ \\
	Radionav. ($\nu$) & $30/50/\text{km}^2$ & $50$  & $2.7$--$5.65$ & $60$ \\
	Radioloca. ($r$)  & $30/50/\text{km}^2$ & $100$  & $2.7$--$5.65$ & $60$ \\
	\bottomrule
\end{tabularx}
\vspace{-5mm}
\end{table}

\vspace{-1mm}
\subsection{Simulation Setup}
\label{subsec:sim_setup}
\vspace{-1mm}
Four services \{$c, s, \nu, r$\} coexist in the same region. General and per-service parameters are listed in Table~\ref{tab:sim_params}; per-service frequency allocations follow \gls{FCC}/\gls{NTIA} assignments in the 2.7--3.7\,GHz federal-band cluster~\cite{ntia_3100_3700,3gpp.38.214}. 
\textcolor{black}{Specifically, 
the shared spectrum pool is centered around  $3.5$\,GHz.}

\textbf{Site-specific ray tracing.} We integrate the BostonTwin\footnote{\href{https://tinyurl.com/2xjn9m43}{https://tinyurl.com/2xjn9m43}} urban digital-twin dataset with the Sionna-RT\footnote{\href{https://nvlabs.github.io/sionna/rt/index.html}{https://nvlabs.github.io/sionna/rt/index.html}} ray-tracing engine~\cite{SS:ModelRealistic24}. A fixed urban tile (\texttt{BOS\_G\_5}) supplies 3D building geometry and real-world cellular base station locations; non-cellular transmitters and receivers are placed by independent homogeneous PPPs within a bounded square aligned with the BostonTwin frame. Antennas are configured via Sionna-RT's \texttt{PlanarArray}, and the \texttt{PathSolver} computes LoS, specular-reflection, and diffraction paths, yielding per-link received power, interference, and \gls{SINR} from complex channel coefficients that combine large- and small-scale effects. The network layout is fixed across Monte-Carlo runs while user locations are resampled.

\textbf{Cellular UE weights.} For long-term proportional fairness~\cite{3gpp.38.214}, UE $u$ has weight
\vspace{-2.5mm}
\begin{equation}
\tilde{w}_u\!=\!\frac{1}{\log_2(1+\widetilde{\gamma}_u)},\;
\widetilde{\gamma}_u\!=\!\frac{\tilde{p}\,\bar{g}_{c,k_c(u),u}}{\tilde{p}\sum_{\ell\neq k_c(u)}\!\!\bar{g}_{c,\ell,u}+N_0 B_{\mathrm{rb}}},
\label{eq:pf_raw_weight}
\end{equation}
with $\tilde{p}\!=\!P_c^{\max}/N_{\mathrm{PRB}}$ and $\bar{g}_{c,k,u}\!=\!\frac{1}{N_{\mathrm{PRB}}T}\sum_{n,t}g_{c,k,u}^{(n)}(t)$. Weights are normalized to unit mean, so cell-edge UEs receive $w_u\!>\!1$ and strong-channel UEs $w_u\!<\!1$. They are computed once and held fixed during optimization, consistent with \gls{3GPP} \gls{NR} \gls{pf} scheduling.

\textbf{Activity and procedure.} Non-cellular transmitters follow predefined patterns with duty cycles $(\delta_s,\delta_\nu,\delta_r)\!=\!(0.2,0.1,0.1)$; cellular BSs are continuously active. Each Monte-Carlo realization samples node and user locations, computes Sionna-RT channels, initializes $\{x,a,p\}$, runs AMRA (Alg.~\ref{alg:alter_all}), and records the weighted cellular throughput, sensing \gls{SNR} margin, and navigation \gls{peb}. All metrics are averaged over 100 realizations.

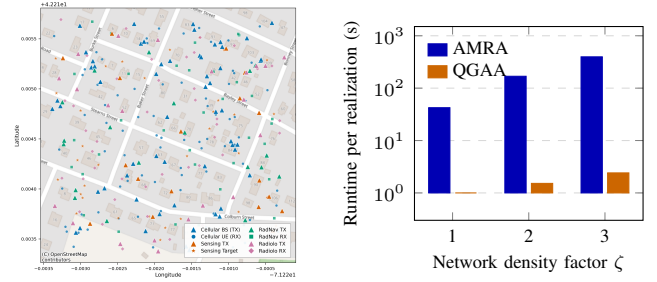
\begin{figure}[t!]
\centering
\begin{subfigure}[b]{0.48\linewidth}
	\centering
	\includegraphics[width=0.9\linewidth]{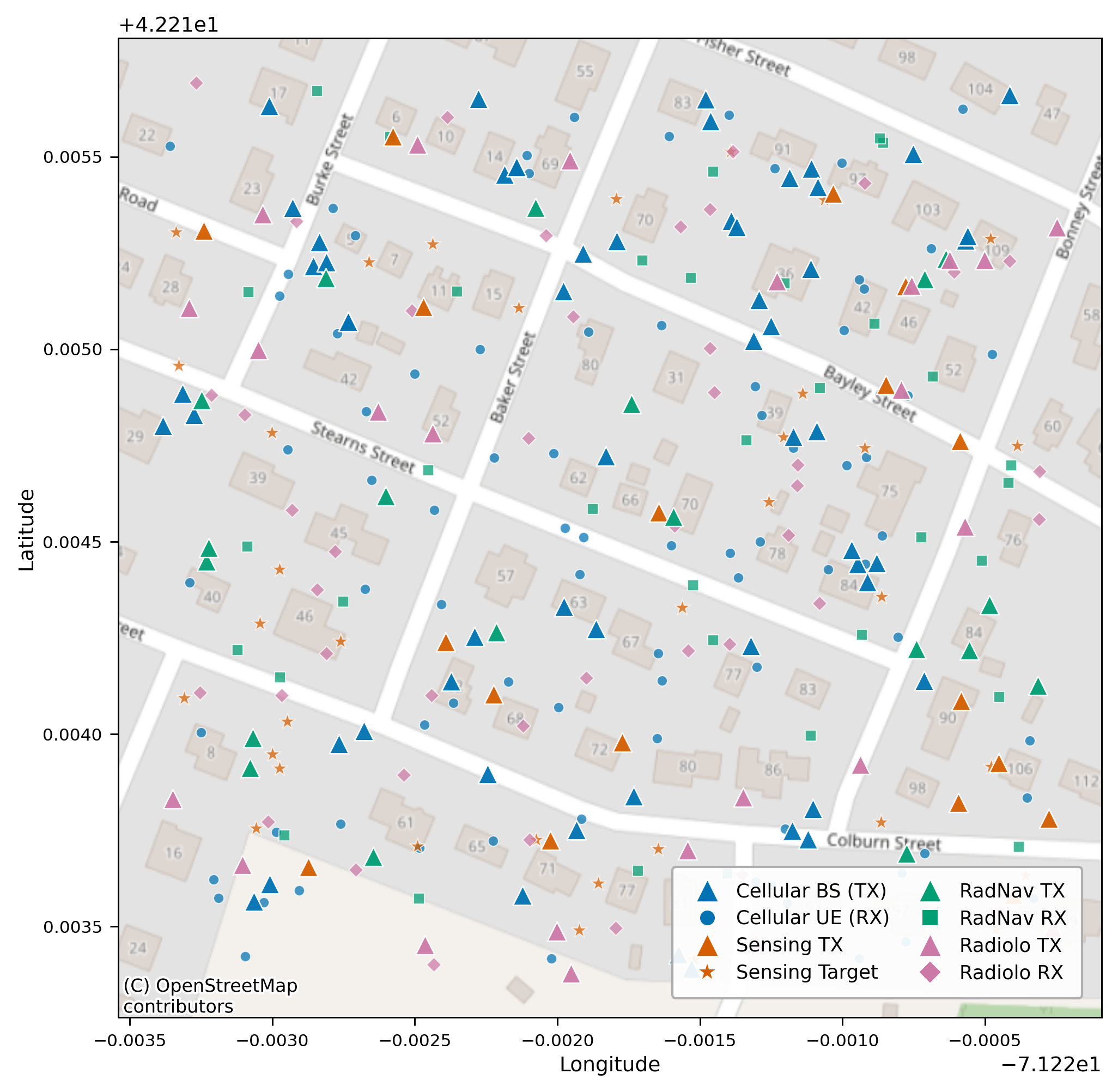}
	\caption{\small Network region ($\zeta\!=\!1$).}
	\label{fig:sionna_test1}
\end{subfigure}\hfill
\begin{subfigure}[b]{0.52\linewidth}
	\centering
	\begin{tikzpicture}
		\begin{axis}[
			width=\linewidth,
			height=0.9\linewidth,
			ybar,
			bar width=8pt,
			ymode=log,
			ymin=0.3, ymax=1500,
			yminorticks=false,
			tick align=inside,
			ylabel={\scriptsize Runtime per realization (s)},
			xlabel={\scriptsize Network density factor $\zeta$},
			symbolic x coords={1,2,3},
			xtick=data,
			xtick align=inside,
			major tick length=3pt,
			xticklabel style={font=\scriptsize},
			yticklabel style={font=\scriptsize},
			ylabel style={font=\scriptsize, yshift=-2pt},
			xlabel style={font=\scriptsize, yshift=2pt},
			legend style={
				font=\scriptsize,
				at={(0.02,0.98)},
				anchor=north west,
				draw=none,
				fill=none,
				row sep=-2pt
			},
			legend cell align=left,
			legend image code/.code={\draw[#1] (0cm,-0.08cm) rectangle (0.25cm,0.08cm);},
			ymajorgrids=true,
			grid style={dashed,gray!30},
			enlarge x limits=0.25,
			]
			\addplot[fill=blue!70!black, draw=blue!70!black] coordinates {
				(1,42.3) (2,168.5) (3,395.7)
			};
			\addlegendentry{AMRA}
			
			\addplot[fill=orange!80!black, draw=orange!80!black] coordinates {
				(1,1.00) (2,1.52) (3,2.41)
			};
			\addlegendentry{QGAA}
		\end{axis}
	\end{tikzpicture}
	\caption{\small Runtime vs.\ network density $\zeta$.}
	\label{fig:complexity_runtime}
\end{subfigure}
\caption{Network deployment region (a) and complexity/runtime comparison of the proposed algorithms (b).}
\label{fig:netw_deploy_sionna}
\vspace{-5mm}
\end{figure}



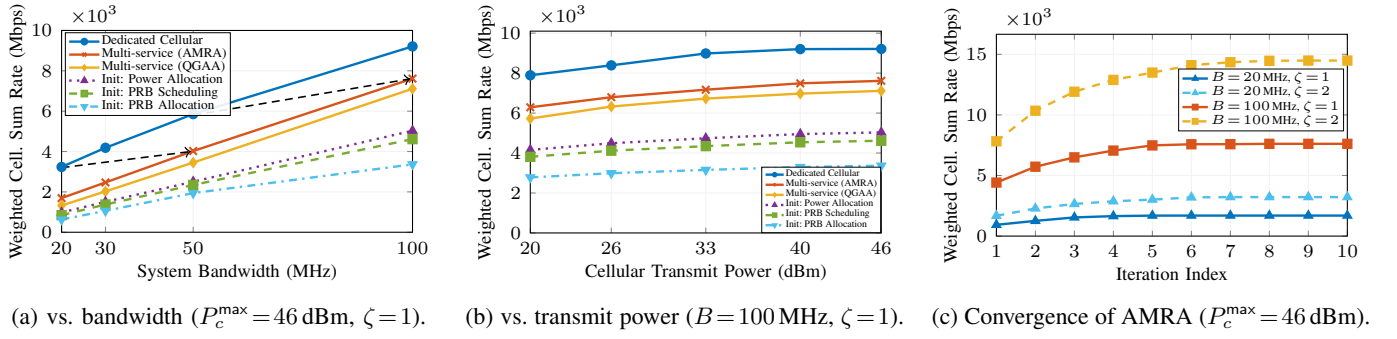
\begin{figure*}[t!]
\centering
\begin{subfigure}{0.32\linewidth}
	\centering
	\begin{tikzpicture}
		\begin{axis}[qosplot, ratey,
			width=0.8\linewidth,
			xlabel={\scriptsize System Bandwidth (MHz)},
			ylabel={\scriptsize Weighted Cell. Sum Rate (Mbps)},
			xmin=20, xmax=100,
			ymin=0,  ymax=10000,
			xtick={20,30,50,100},
			major tick length=2pt,
			legend style={
				at={(0.01,0.99)}, anchor=north west,
				font=\fontsize{5.0}{5.5}\selectfont,
				inner xsep=1pt, inner ysep=0.5pt, row sep=-0.8pt,
				nodes={inner sep=0.4pt},
			},
			legend image post style={scale=0.5},
			legend image code/.code={%
				\draw[#1, mark=none]
				plot coordinates {(0cm,0cm) (0.8cm,0cm)};%
				\draw[#1, only marks, mark size=2pt]
				plot coordinates {(0.4cm,0cm)};%
			},
			]
			\addplot[matBlue,   mark=*,   mark options={fill=matBlue}]
			coordinates {
				(20,3236)
				(30,4185)
				(50,5854)
				(100,9210)};
			\addlegendentry{Dedicated Cellular}
			\addplot[matOrange, mark=x,   mark options={solid, line width=0.8pt, mark size=2pt}]
			coordinates {
				(20,1692)
				(30,2468)
				(50,4025)
				(100,7613)};
			\addlegendentry{Multi-service (AMRA)}
			\addplot[matYellow, mark=diamond*, mark options={fill=matYellow, mark size=1.5pt}]
			coordinates {
				(20,1325)
				(30,2024)
				(50,3452)
				(100,7112)};
			\addlegendentry{Multi-service (QGAA)}
			\addplot[matPurple, dotted,        mark=triangle*,
			mark options={solid,fill=matPurple}]
			coordinates {
				(20,979)
				(30,1508)
				(50,2504)
				(100,5032)};
			\addlegendentry{Init: Power Allocation}
			\addplot[matGreen,  dashed,        mark=square*,
			mark options={solid,fill=matGreen}]
			coordinates {
				(20,846)
				(30,1375)
				(50,2343)
				(100,4614)};
			\addlegendentry{Init: PRB Scheduling}
			\addplot[matCyan,   dashdotted,    mark=triangle*,
			mark options={solid, rotate=180,
				fill=matCyan}]
			coordinates {
				(20,631)
				(30,1061)
				(50,1940)
				(100,3365)};
			\addlegendentry{Init: PRB Allocation}
			
			\draw[->, thick, black, line width=0.6pt, densely dashed]
			(axis cs:20, 3200) -- (axis cs:49.5, 4000);
			
			\draw[->, thick, black, line width=0.6pt, densely dashed]
			(axis cs:50, 5800) -- (axis cs:99.5, 7600);
			
		\end{axis}
	\end{tikzpicture}
	\caption{vs.\ bandwidth ($P_c^{\sf max}\!=\!46$\,dBm, $\zeta\!=\!1$).}
	\label{fig:weight_rate_bw_compare_theo_sion}
\end{subfigure}\hfill
\begin{subfigure}{0.32\linewidth}
	\centering
	\begin{tikzpicture}
		\begin{axis}[qosplot, ratey,
			width=0.8\linewidth,
			xlabel={\scriptsize Cellular Transmit Power (dBm)},
			ylabel={\scriptsize Weighted Cell. Sum Rate (Mbps)},
			xmin=20, xmax=46,
			ymin=0, ymax=10100,
			xtick={20,26,33,40,46},
			legend style={
				at={(0.999,0.00)}, anchor=south east,
				font=\fontsize{3.7}{5}\selectfont,
				inner xsep=1pt, inner ysep=0.5pt, row sep=-0.8pt,
				nodes={inner sep=0.3pt},
			},
			legend image post style={scale=0.3},
			legend image code/.code={%
				\draw[#1, mark=none]
				plot coordinates {(0cm,0cm) (0.8cm,0cm)};%
				\draw[#1, only marks, mark size=2pt]
				plot coordinates {(0.4cm,0cm)};%
			},
			]
			\addplot[matBlue,   mark=*,        mark options={fill=matBlue}]
			coordinates {
				(20,7892)
				(26,8389)
				(33,8983)
				(40,9201)
				(46,9210)};
			\addlegendentry{Dedicated Cellular}
			\addplot[matOrange, mark=x,  mark options={solid, line width=0.8pt, mark size=2pt}]
			coordinates {
				(20,6289)
				(26,6793)
				(33,7170)
				(40,7488)
				(46,7613)};
			\addlegendentry{Multi-service (AMRA)}
			\addplot[matYellow, mark=diamond*, mark options={fill=matYellow, mark size=1.5pt}]
			coordinates {
				(20,5725)
				(26,6326)
				(33,6725)
				(40,6975)
				(46,7112)};
			\addlegendentry{Multi-service (QGAA)}
			\addplot[matPurple, dotted,        mark=triangle*,
			mark options={solid,fill=matPurple}]
			coordinates {
				(20,4157)
				(26,4490)
				(33,4739)
				(40,4949)
				(46,5032)};
			\addlegendentry{Init: Power Allocation}
			\addplot[matGreen,  dashed,        mark=square*,
			mark options={solid,fill=matGreen}]
			coordinates {
				(20,3811)
				(26,4117)
				(33,4345)
				(40,4538)
				(46,4614)};
			\addlegendentry{Init: PRB Scheduling}
			\addplot[matCyan,   dashdotted,    mark=triangle*,
			mark options={solid, rotate=180,
				fill=matCyan}]
			coordinates {
				(20,2780)
				(26,2989)
				(33,3155)
				(40,3295)
				(46,3365)};
			\addlegendentry{Init: PRB Allocation}
		\end{axis}
	\end{tikzpicture}
	\caption{vs.\ transmit power ($B\!=\!100$\,MHz, $\zeta\!=\!1$).}
	\label{fig:weight_rate_tx_sion}
\end{subfigure}\hfill
\begin{subfigure}{0.32\linewidth}
	\centering
	\begin{tikzpicture}
		\begin{axis}[qosplot, ratey,
			width=0.8\linewidth,
			xlabel={\scriptsize Iteration Index},
			ylabel={\scriptsize Weighted Cell. Sum Rate (Mbps)},
			xmin=1, xmax=10,
			ymin=0, ymax=16650,           
			xtick={1,2,3,4,5,6,7,8,9,10},
			legend style={
				at={(0.99,0.52)}, anchor=south east,
				font=\fontsize{4.5}{5.5}\selectfont,
				inner xsep=1pt, inner ysep=0.5pt, row sep=-0.8pt,
				nodes={inner sep=0.6pt},
			},
			legend image post style={scale=0.42},
			legend image code/.code={%
				\draw[#1, mark=none]
				plot coordinates {(0cm,0cm) (0.8cm,0cm)};%
				\draw[#1, only marks, mark size=2pt]
				plot coordinates {(0.4cm,0cm)};%
			},
			]
			\addplot[matBlue, mark=triangle*, mark options={solid,fill=matBlue}]
			coordinates {
				(1,931)
				(2,1255)
				(3,1541)
				(4,1648)
				(5,1689)
				(6,1692)
				(7,1692)
				(8,1692)
				(9,1692)
				(10,1692)};
			\addlegendentry{$B\!=\!20$\,MHz, $\zeta\!=\!1$}
			\addplot[matCyan, dashed,
			mark=triangle*, mark options={solid,fill=matCyan}]
			coordinates {
				(1,1673)
				(2,2281)
				(3,2650)
				(4,2873)
				(5,3009)
				(6,3191)
				(7,3218)
				(8,3218)
				(9,3218)
				(10,3218)};
			\addlegendentry{$B\!=\!20$\,MHz, $\zeta\!=\!2$}
			\addplot[matOrange,
			mark=square*, mark options={fill=matOrange}]
			coordinates {
				(1,4416)
				(2,5724)
				(3,6497)
				(4,7054)
				(5,7483)
				(6,7577)
				(7,7583)
				(8,7613)
				(9,7613)
				(10,7613)};
			\addlegendentry{$B\!=\!100$\,MHz, $\zeta\!=\!1$}
			\addplot[matYellow, dashed,
			mark=square*, mark options={solid,fill=matYellow}]
			coordinates {
				(1,7820)
				(2,10344)
				(3,11912)
				(4,12886)
				(5,13490)
				(6,14099)
				(7,14344)
				(8,14462)
				(9,14482)
				(10,14482)};
			\addlegendentry{$B\!=\!100$\,MHz, $\zeta\!=\!2$}
		\end{axis}
	\end{tikzpicture}
	\caption{Convergence of AMRA ($P_c^{\sf max}\!=\!46$\,dBm).}
	\label{fig:weight_converge_sion}
\end{subfigure}

\caption{Weighted cellular sum-rate: dedicated cellular vs.\ multi-service AMRA and QGAA.}
\label{fig:weight_rate_bw_tx}
\vspace{-4.5mm}
\end{figure*}


\begin{figure*}[t!]
\centering
\begin{subfigure}{0.32\linewidth}
	\centering
	\begin{tikzpicture}
		\begin{axis}[qosplot, ratey,
			width=0.8\linewidth,
			xlabel={\scriptsize Multiplied BS/UE Density $\zeta$},
			ylabel={\scriptsize Weighted Cell. Sum Rate (Mbps)},
			xmin=1, xmax=3,
			ymin=0,   ymax=24800,
			xtick={1.0,1.5,2.0,2.5,3.0},
			xticklabels={$1.0$,$1.5$,$2.0$,$2.5$,$3.0$},
			major tick length=2pt,
			legend pos=south east,
			legend style={
				font=\fontsize{6.0}{5.5}\selectfont,
				inner xsep=1pt, inner ysep=0.5pt, row sep=-0.8pt,
				nodes={inner sep=0.5pt},
			},
			legend image post style={scale=0.5},
			legend image code/.code={%
				\draw[#1, mark=none]
				plot coordinates {(0cm,0cm) (0.8cm,0cm)};%
				\draw[#1, only marks, mark size=2pt]
				plot coordinates {(0.4cm,0cm)};%
			},
			]
			legend style={
				inner xsep=1pt, inner ysep=1pt, row sep=-3pt,
				draw=none, fill=none,
				nodes={inner sep=1pt, scale=0.9},
			},
			\addplot[matBlue,   mark=*,         mark options={fill=matBlue}]
			coordinates {
				(1.0,9210)
				(1.5,12552)
				(2.0,16025)
				(2.5,19361)
				(3.0,22515)};
			\addlegendentry{Dedicated Cellular}
			\addplot[matOrange, mark=square*,   mark options={fill=matOrange}]
			coordinates {
				(1.0,7613)
				(1.5,11051)
				(2.0,14482)
				(2.5,17822)
				(3.0,21088)};
			\addlegendentry{Multi-service (AMRA)}
			\addplot[matYellow, mark=triangle*, mark options={solid,fill=matYellow}, dashed]
			coordinates {
				(1.0,7112)
				(1.5,10546)
				(2.0,14120)
				(2.5,17432)
				(3.0,20526)};
			\addlegendentry{Multi-service (QGAA)}
		\end{axis}
	\end{tikzpicture}
	\caption{vs.\ network density scaling factor $\zeta$.}
	\label{fig:weight_rate_dens_sion}
\end{subfigure}\hfill
\begin{subfigure}{0.33\linewidth}
	\centering
	\begin{tikzpicture}
		\begin{axis}[qosplot,
			axis y line*=right, axis x line*=top,
			xticklabels=\empty,  
			xtick={1,2,3,4,5,6,7},
			xmin=0.5, xmax=7.5,
			ymin=0,   ymax=453.6,
			ylabel={\scriptsize Number of PRBs},
			ytick={0,100,200,300,400},
			major tick length=2pt,
			grid=none,
			]
			\addplot+[ybar, bar width=5pt, mark=none,
			draw=matOrange, line width=0.4pt,
			pattern={north east lines}, pattern color=matOrange]
			coordinates {
				(0.825,68)
				(1.825,90)
				(2.825,113)
				(3.825,157)
				(4.825,269)
				(5.825,324)
				(6.825,324)};
			\addplot+[ybar, bar width=5pt, mark=none,
			draw=matPurple, line width=0.4pt,
			fill=matPurple,
			pattern color=matPurple]
			coordinates {
				(1.175,34)
				(2.175,45)
				(3.175,56)
				(4.175,79)
				(5.175,134)
				(6.175,162)
				(7.175,162)};
		\end{axis}
		\begin{axis}[qosplot, ratey,
			axis y line*=left, axis x line*=bottom,
			xlabel={\scriptsize Sensing SNR Threshold $\gamma^{\sf sen}$ (dB)},
			ylabel={\scriptsize Weighted Cell. Sum Rate (Mbps)},
			xmin=0.5, xmax=7.5,
			ymin=1523, ymax=11052,
			xtick={1,2,3,4,5,6,7},
			xticklabels={0,3,4.8,7,10,13,15},
			major tick length=2pt,
			legend style={
				at={(0.01,0.42)}, anchor=west,
				font=\fontsize{4.0}{5.5}\selectfont,
				inner xsep=1pt, inner ysep=0.5pt, row sep=-0.8pt,
				nodes={inner sep=0.5pt},
			},
			legend image post style={scale=0.5},
			legend image code/.code={%
				\draw[#1, mark=none]
				plot coordinates {(0cm,0cm) (0.8cm,0cm)};%
				\draw[#1, only marks, mark size=2pt]
				plot coordinates {(0.4cm,0cm)};%
			},
			]
			\addplot[matBlue,  mark=*,  mark options={fill=matBlue}]
			coordinates {
				(1,9210)
				(2,9210)
				(3,9210)
				(4,9210)
				(5,9210)
				(6,9210)
				(7,9210)};
			\addlegendentry{Dedicated Cellular}
			\addplot[matOrange,   mark=square*,   mark options={fill=matOrange}]
			coordinates {
				(1,8222)
				(2,7841)
				(3,7613)
				(4,6928)
				(5,5443)
				(6,3197)
				(7,2246)};
			\addlegendentry{Multi-service (AMRA)}
			\addplot[matYellow, mark=triangle*, mark options={solid,fill=matYellow}, dashed]
			coordinates {
				(1,7651)
				(2,7270)
				(3,7042)
				(4,6357)
				(5,4872)
				(6,2665)
				(7,1751)};
			\addlegendentry{Multi-service (QGAA)}
			\addlegendimage{area legend, draw=matOrange, line width=0.4pt,
				pattern={north east lines}, pattern color=matOrange}
			\addlegendentry{PRBs: Sensing}
			\addlegendimage{area legend, draw=matPurple, line width=0.4pt,
				fill={matPurple}, pattern color=matPurple}
			\addlegendentry{PRBs: Radiolocation}
			\addplot[only marks, mark=pentagon*, mark size=2pt,
			mark options={fill=starFill, draw=black, line width=0.4pt},
			forget plot]
			coordinates {(3,7613)};
		\end{axis}
	\end{tikzpicture}
	\caption{vs.\ sensing SNR threshold $\gamma^{\sf sen}$ ($\zeta\!=\!1$).}
	\label{fig:weight_rate_sens_sion}
\end{subfigure}\hfill
\begin{subfigure}{0.33\linewidth}
	\centering
	\begin{tikzpicture}
		\begin{axis}[qosplot,
			axis y line*=right, axis x line*=top,
			xtick={1,1.5,2,2.83,5,10},
			xticklabels=\empty,  
			x dir=reverse,
			xmin=0.8, xmax=10.5,
			ymin=0,   ymax=512.5,
			ylabel={\scriptsize Number of PRBs},
			ytick={0,100,200,300,400,500},
			major tick length=2pt,
			grid=none,
			]
			\addplot+[ybar, bar width=5pt, draw=navGreen, mark=none,
			fill=navGreen, fill opacity=0.55]
			coordinates {
				(10,14)
				(5,26)
				(2.83,60)
				(2,110)
				(1.5,188)
				(1,410)};
		\end{axis}
		\begin{axis}[qosplot, ratey,
			axis y line*=left, axis x line*=bottom,
			x dir=reverse,
			xlabel={\scriptsize Navigation PEB Threshold $\epsilon_{\max}$ (m)},
			ylabel={\scriptsize Weighted Cell. Sum Rate (Mbps)},
			xmin=0.8, xmax=10.5,
			ymin=0,   ymax=11052,
			major tick length=2pt,
			xtick={1,1.5,2,2.83,5,10},
			xticklabel style={/pgf/number format/.cd, fixed, precision=2,
				/tikz/.cd},
			legend style={
				at={(0.03,0.32)}, anchor=west,
				font=\fontsize{6.0}{5.5}\selectfont,
				inner xsep=1pt, inner ysep=0.5pt, row sep=-0.8pt,
				nodes={inner sep=0.5pt},
			},
			legend image post style={scale=0.5},
			legend image code/.code={%
				\draw[#1, mark=none]
				plot coordinates {(0cm,0cm) (0.8cm,0cm)};%
				\draw[#1, only marks, mark size=2pt]
				plot coordinates {(0.4cm,0cm)};%
			},
			]
			\addplot[matBlue,  mark=*,         mark options={fill=matBlue}]
			coordinates {
				(10,9210)
				(5,9210)
				(2.83,9210)
				(2,9210)
				(1.5,9210)
				(1,9210)};
			\addlegendentry{Dedicated Cellular}
			\addplot[matOrange,   mark=square*,   mark options={fill=matOrange}]
			coordinates {
				(10,8603)
				(5,8032)
				(2.83,7613)
				(2,6699)
				(1.5,4872)
				(1,876)};
			\addlegendentry{Multi-service (AMRA)}
			\addplot[matYellow, mark=triangle*, mark options={solid,fill=matYellow}, dashed]
			coordinates {
				(10,7956)
				(5,7423)
				(2.83,7042)
				(2,6090)
				(1.5,4263)
				(1,571)};
			\addlegendentry{Multi-service (QGAA)}
			\addlegendimage{area legend, fill=navGreen, fill opacity=0.55, draw=navGreen}
			\addlegendentry{PRBs: Radionavigation}
			\addplot[only marks, mark=pentagon*, mark size=2pt,
			mark options={fill=starFill, draw=black, line width=0.4pt},
			forget plot]
			coordinates {(2.83,7613)};
		\end{axis}
	\end{tikzpicture}
	\caption{vs.\ navigation PEB threshold $\epsilon_{\max}$ ($\zeta\!=\!1$).}
	\label{fig:weight_rate_nav_sion}
\end{subfigure}
\caption{Weighted cellular sum-rate vs.\ network density and \gls{QoS}
	thresholds ($B\!=\!100$\,MHz, $P_c^{\sf max}\!=\!46$\,dBm).}
\label{fig:weight_rate_qos}
\end{figure*}
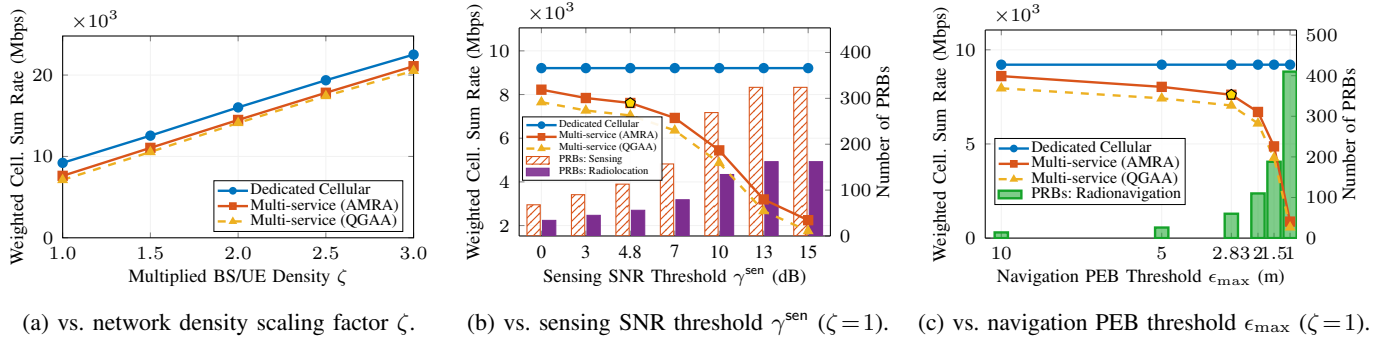

\vspace{-2mm}
\subsection{Results and Discussion}
\vspace{-1mm}

\textcolor{black}{
Fig.~\ref{fig:netw_deploy_sionna}(a) shows the BostonTwin deployment region; Fig.~\ref{fig:netw_deploy_sionna}(b) reports the runtime of AMRA and QGAA versus network density factor $\zeta$. 
As predicted by the complexity analysis in~\eqref{eq:complexity_overall}, AMRA exhibits steep cubic-like growth in $\zeta$, whereas QGAA scales near-linearly and stays around $1$\,s at $\zeta\!=\!1$, confirming its suitability for real-time deployment as an \gls{ORAN} Near-RT \gls{RIC} xApp.}

\textbf{Throughput vs.\ bandwidth and power.} Figs.\,\ref{fig:weight_rate_bw_compare_theo_sion}
and~\ref{fig:weight_rate_tx_sion} report the weighted cellular sum-rate against bandwidth and cellular transmit power. Throughput grows monotonically with both, and AMRA consistently outperforms QGAA across operating points, with the gap reflecting the cost of preserving sensing and navigation \gls{QoS}. 
\textcolor{black}{A central observation, highlighted by the dashed arrows in Fig.\,\ref{fig:weight_rate_bw_compare_theo_sion}, is the \emph{cross-bandwidth crossover}: AMRA at $B\!=\!50$\,MHz exceeds dedicated cellular at $B\!=\!20$\,MHz, and AMRA at $B\!=\!100$\,MHz exceeds dedicated at $B\!=\!50$\,MHz; QGAA exhibits the same behavior. 
Faced with a choice between a narrow exclusive cellular allocation and a wider shared one, sharing yields strictly more cellular capacity \emph{and} serves sensing, navigation, and radar, the central motivation for coordinated multi-service sharing.}
Moreover, Fig.~\ref{fig:weight_converge_sion} confirms that the weighted sum-rate increases monotonically and converges within a few outer AO iterations across bandwidth and density settings, demonstrating stability and scalability under realistic propagation.


\textbf{Network density and \gls{QoS} sensitivity.}
Fig.~\ref{fig:weight_rate_qos} evaluates the weighted cellular sum-rate under varying network density and \gls{QoS} constraints. Fig.~\ref{fig:weight_rate_dens_sion} shows all schemes scaling near-linearly with density factor $\zeta$, with multi-service schemes tracking the dedicated baseline closely, and coexistence overhead does not compound with densification. Tightening the sensing \gls{SNR} threshold $\gamma^{\text{sen}}$  (Fig.~\ref{fig:weight_rate_sens_sion}) increases \gls{prb} allocation to sensing, eroding the cellular sum-rate near-linearly. In contrast, tightening the navigation \gls{peb} threshold $\epsilon_{\max}$   (Fig.~\ref{fig:weight_rate_nav_sion}) induces a \emph{convex} collapse driven by the non-linear cost of sub-meter positioning accuracy. This asymmetry persists across channel models, indicating a structural property of the \gls{QoS} constraints~\eqref{C:P2_sensing}--\eqref{C:P2_nav} rather than an artifact of analytical shadowing.


\vspace{-3.5mm}
\section{Conclusion}
\label{sec:conclusion}
\vspace{-1mm}

We presented the first unified resource-allocation framework for coordinated spectrum sharing across cellular, \gls{RF} sensing, radionavigation, and radiolocation services, formulated as an \gls{MINLP} and decomposed into \gls{prb} assignment, scheduling, and SCA-based power
allocation, complemented by a low-complexity \gls{QoS}-aware greedy heuristic.
Evaluations on the BostonTwin urban digital twin demonstrate substantial cellular-throughput gains while strictly satisfying sensing and navigation \gls{QoS} constraints. 
The framework maps naturally onto an \gls{ORAN} architecture, with AMRA as a
Non-RT \gls{RIC} rApp issuing A1 policies to a Near-RT \gls{RIC} xApp executing QGAA.
Future work targets learning-based and distributed implementations, channel-uncertainty robustness, chance-constrained \gls{peb} formulations, and O-RAN testbed validation.


\vspace{-3.5mm}
\footnotesize{
\bibliographystyle{IEEEtran}
\bibliography{ref_26}
}

\end{document}